\documentclass[journal,12pt,onecolumn,draftclsnofoot,a4paper,]{IEEEtran}

\usepackage{cite}
\usepackage[T1]{fontenc}
\usepackage[latin1]{inputenc}
\usepackage{graphicx}
\usepackage{theorem}
\usepackage{amsmath}
\usepackage{amsfonts}
\ifCLASSOPTIONcompsoc
\usepackage[caption=false,font=normalsize,labelfon
t=sf,textfont=sf]{subfig}
\else
\usepackage[caption=false,font=footnotesize]{subfi
g}
\fi
\usepackage[acronym,shortcuts]{glossaries}
\usepackage{color}
\usepackage{url}
\usepackage{enumerate}
\usepackage{comment}
\usepackage{tikz}
\usetikzlibrary{shapes,arrows}
\usepackage{cleveref}
\usepackage{multirow}

\usepackage{algorithm}
\usepackage[noend]{algpseudocode}
\algdef{SE}[DOWHILE]{Do}{doWhile}{\algorithmicdo}[1]{\algorithmicwhile\ #1}%

\newacronym{QC}{QC}{quasi-cyclic}
\newacronym{QC-LDPC}{QC-LDPC}{quasi-cyclic low-density parity-check}
\newacronym{LDPC}{LDPC}{low-density parity-check}
\newacronym{LDPCC}{LDPCC}{low-density parity-check convolutional}
\newacronym{AC-LDPC}{AC-LDPC}{array convolutional low-density parity-check}
\newacronym{PDC-LDPC}{PDC-LDPC}{progressive differences convolutional low-density parity-check}
\newacronym{SC-LDPC}{SC-LDPC}{spatially coupled low-density parity-check}
\newacronym{SC-LDPC-CCs}{SC-LDPCC}{spatially coupled LDPC code}
\newacronym{SC-LDPC-CC}{SC-LDPCC}{spatially coupled LDPC code}
\newacronym{AWGN}{AWGN}{additive white Gaussian noise}
\newacronym{BER}{BER}{bit error rate}
\newacronym{FER}{FER}{codeword error rate}
\newacronym{TUB}{TUB}{truncated union bound}
\newacronym{BPSK}{BPSK}{binary phase shift keying}
\newacronym{SPA-LLR}{SPA-LLR}{sum-product algorithm with log-likelihood ratios}
\newacronym{RTI}{RTI}{regular time-invariant}
\newacronym{RTI-LDPCC}{RTI-LDPCC}{regular time-invariant low-density parity-check convolutional}
\newacronym{AS}{AS}{absorbing set}
\newacronym{CPMs}{CPMs}{circulant permutation matrices}
\newacronym{SW}{SW}{sliding window}

{\theorembodyfont{\rmfamily}
   }
{\theorembodyfont{\rmfamily}
   }
{\theorembodyfont{\rmfamily}
   \newtheorem{Lem}{{\textbf Lemma}}}
{\theorembodyfont{\rmfamily}
   \newtheorem{Cor}{{\textbf Corollary}}}
{\theorembodyfont{\rmfamily}
   }
{\theorembodyfont{\rmfamily}
   \newtheorem{Exa}{{\textbf Example}}}
{\theorembodyfont{\rmfamily}
   }
 
\def\HH{\mathbf{H}}

\def\PP{\mathbf{P}}
\def\BB{\mathbf{B}}
\def\bb{\mathbf{b}}

\def\0{\mathbf{0}}

\newcommand{\mb}[1]{\textcolor{red}{#1}}

\tikzstyle{decision} = [diamond, draw, fill=green!20, 
    text width=4.5em, text badly centered, node distance=3cm, inner sep=0pt]
\tikzstyle{block} = [rectangle, draw, fill=blue!20, 
    text width=10em, text centered, rounded corners, minimum height=4em]
\tikzstyle{line} = [draw, -latex']
\tikzstyle{cloud} = [draw, ellipse, fill=red!20, node distance=3cm, minimum height=2em]

\makeatletter
\newcommand{\tpmod}[1]{{\@displayfalse\pmod{#1}}}
\newcommand{\lr}{\hspace{2mm}\textbf{to}\hspace{2mm}}
\makeatother

\begin{document}

\title{Efficient Search and Elimination of Harmful Objects in Optimized QC SC-LDPC Codes}

\author{\IEEEauthorblockN{Massimo Battaglioni\IEEEauthorrefmark{1},
Franco Chiaraluce\IEEEauthorrefmark{1}, Marco Baldi\IEEEauthorrefmark{1}, and David Mitchell\IEEEauthorrefmark{2}}
\IEEEauthorblockA{\IEEEauthorrefmark{1}Dipartimento di Ingegneria dell'Informazione, Universit\`a Politecnica delle Marche, Ancona, Italy\\
Email: \{m.battaglioni, f.chiaraluce, m.baldi\}@staff.univpm.it}\\
\IEEEauthorblockA{\IEEEauthorrefmark{2}Klipsch School of Electrical and Computer Engineering, New Mexico State University, Las Cruces, NM 88011\\
Email: dgmm@nmsu.edu}
}

\maketitle
\begin{abstract}
The error correction performance of low-density parity-check (LDPC) codes under iterative message-passing decoding is degraded by the presence of certain harmful objects existing in their Tanner graph representation. Depending on the context, such harmful objects are known as stopping sets, trapping sets, absorbing sets, or pseudocodewords.
In this paper, we propose a general procedure based on \emph{edge spreading} that enables the design of quasi-cyclic (QC) spatially coupled
low-density parity-check codes (SC-LDPCCs) that are derived from QC-LDPC block codes and possess a significantly reduced multiplicity of harmful objects with respect to the original QC-LDPC block code. This procedure relies on a novel algorithm that greedily spans the search space of potential candidates to reduce the multiplicity of the target harmful object(s) in the Tanner graph. The effectiveness of the method we propose is validated via examples and numerical computer simulations.
\end{abstract}


\section{Introduction}

\Ac{LDPC} block codes were first introduced by Gallager \cite{Gallager} and have attracted significant interest over time due to their capacity-approaching performance. The convolutional counterpart of \ac{LDPC} block codes, called LDPC convolutional codes or \acp{SC-LDPC-CCs}, were first proposed in \cite{Felstrom1999}. Recent studies have shown that SC-LDPCCs are able to achieve the capacity of memoryless binary-input output symmetric channels under iterative decoding 
based on belief propagation \cite{Lentmaier2010,Kudekar2011}.


It is well known that iterative algorithms used for decoding \ac{LDPC} codes can get trapped in certain error patterns that arise due to structural imperfections in the code's Tanner graph. These objects may cause a severe degradation of the error correction performance, especially in the high signal-to-noise ratio region (\textit{error-floor} region).
These harmful objects depend on the considered channel and the type of decoding algorithm in use. The concept of \textit{stopping set} was introduced in \cite{Di2002}, where the failures of iterative algorithms over the binary erasure channel are characterized. More complex channels, like the \ac{AWGN} channel, require the definition of more subtle harmful objects. The first work in this direction is \cite{Richardson2001}, where \textit{trapping sets} are defined. A particularly harmful subclass of trapping sets, called \textit{absorbing sets}, were shown to be stable under bit-flipping iterative decoders \cite{Dolecek2010}. 
It was shown in \cite{Hashemi2014, Hashemi2016} that starting from a cycle, or from a cluster of cycles, in the Tanner graph of a regular or irregular \ac{LDPC} code, any trapping set can be obtained by means of some graph expansion technique.

\Acp{SC-LDPC-CCs} can be designed starting from \ac{LDPC} block codes via an \textit{edge spreading} procedure \cite{Mitchell2015}, that is a generalization of the unwrapping techniques introduced in \cite{Felstrom1999,Tanner2004}. 
Clearly, the harmful objects of the \acp{SC-LDPC-CCs} arise from related objects in the underlying \ac{LDPC} block codes, and their multiplicity depends on the adopted edge spreading method. Some efforts have been devoted to the graph optimization from an absorbing set standpoint of array-based \acp{SC-LDPC-CCs} \cite{Mitchell2014,Amiri2016,Mitchell2017,Beemer2016,Beemer2017,Esfahanizadeh2017, Esfahanizadeh2019}. These approaches have been restricted to certain code structures and harmful objects to enable a feasible search.
Furthermore, most of these previous works have the limitation of excluding a priori many possible solutions of the problem, in order to reduce the search space.
Moreover, as shown in \cite{Mitchell2017,Esfahanizadeh2017}, the multiplicity of harmful objects can be significantly reduced by increasing the memory of SC-LDPCCs. However, the computational complexity of previous approaches limits their viability to small memories. To the best of the authors' knowledge, a general scheme enabling the construction of optimized quasi-cyclic SC-LDPCCs (QC-SC-LDPCCs) (with respect to minimization of harmful objects) from QC-LDPC block codes with large memories is missing from the literature.

The objective of this paper is to propose an algorithm that, given any QC-LDPC block code exploits a smart strategy to construct an optimized QC-SC-LDPCC by performing a greedy search over all candidates. This search attempts to minimize the multiplicity of the most harmful object (or combinations of objects) for the given channel and decoding algorithm. The effectiveness of the proposed algorithm is demonstrated for several exemplary code constructions with varying code memories via enumeration of the target harmful objects and numerical computer simulations.

The paper is organized as follows. In Section \ref{sec:TICodes} we introduce the notation used throughout the paper and basic notions of QC-LDPC block codes and \acp{SC-LDPC-CCs} derived from them. In Section \ref{sec:edgesp} we focus on edge spreading matrices and the corresponding cycle properties. In Section \ref{sec:algo} we describe the algorithm we propose. In Section \ref{sec:perf} we provide some examples and assess their error rate performance. Finally, in Section \ref{sec:conc} we draw some conclusions.

\section{Definitions and notation \label{sec:TICodes}}

In this section we first introduce the notation for QC-LDPC codes and describe the edge spreading procedure to obtain QC-SC-LDPCCs from QC-LDPC block codes.

\subsection{QC-LDPC codes \label{subsec:qc}}

Let us consider a QC-LDPC block code, in which the parity-check matrix $\HH$ is an $m \times n$ array of $N \times N$ \ac{CPMs} or all-zero matrices. We denote these matrices as $\mathbf{I}(p_{i,j})$, $0 \leq i \leq m - 1$, $0 \leq j \leq n - 1$, while $N$ is the \textit{lifting degree} of the code and $p_{i,j}\in \{-\infty, 0, 1, \ldots, N-1 \}$. When $0 \leq p_{i,j} \leq N - 1$, $\mathbf{I}(p_{i,j})$ is obtained from the identity matrix through a cyclic shift of its rows to the left/right by $p_{i,j}$ positions. We instead conventionally denote the all zero matrix by $\mathbf{I}(-\infty)$. The code length is $L=nN$. The {\em exponent matrix} of the code is the $m \times n$ matrix $\mathbf{P}$ having the values $p_{i,j}$ as its entries.


We associate a Tanner graph $\mathcal{G}(\HH)$ to any parity-check matrix $\HH$ as follows:
\begin{itemize}
\item any column of $\HH$ corresponds to a variable node;
\item any row of $\HH$ corresponds to a check node;
\item there is an edge between the $i$th check node and the $j$th variable node if and only if the $(i,j)$th entry of $\HH$ is $1$.
\end{itemize}
The set of $L$ variable nodes is denoted as $\mathcal{V}$ and the set of $mN$ check nodes is denoted as $\mathcal{P}$. The set of edges is denoted as $E$. Thus, we can express $\mathcal{G}(\HH)$ as $\mathcal{G}(\mathcal{V} \cup \mathcal{P},E)$.
Let us consider the subgraph induced by a subset $\mathcal{D}$ of $\mathcal{V}$. We define $\mathcal{E}(\mathcal{D})$ and $\mathcal{O}(\mathcal{D})$ as the set of neighboring check nodes with even and odd degree in such subgraph, respectively. The \textit{girth} of $\mathcal{G}(\HH)$, noted by $g$, is the length of the shortest cycle in the graph.

An $(a,b)$ \textit{\ac{AS}} is a subset  $\mathcal{D}$ of $\mathcal{V}$ of size $a > 0$, with  $\mathcal{O}(\mathcal{D})$ of size $b\geq 0$ and with the property that each
variable node in $\mathcal{D}$ has strictly fewer neighbors in $\mathcal{O}(\mathcal{D})$ than in $\mathcal{C} \setminus \mathcal{O}(\mathcal{D})$. 
We say that an $(a, b)$ \ac{AS} $\mathcal{D}$ is an $(a, b)$
\textit{fully AS (FAS)} if, in addition, all nodes in $\mathcal{V} \setminus \mathcal{D}$ have
strictly more neighbors in $\mathcal{C} \setminus \mathcal{O}(\mathcal{D})$ than in $\mathcal{O}(\mathcal{D})$.

For a QC-LDPC code, a necessary and sufficient condition for the existence of a cycle of length $2k$ 
in $\mathcal{G}(\HH)$ is \cite{Fossorier2004}
\begin{equation}
\sum_{i=0}^{k-1} \left( p_{m_{i},n_{i}} - p_{m_{i},n_{i+1}} \right) = 0  \mod N , 
\label{fore}
\end{equation}   
where $n_{k}=n_{0}$, $m_{i} \neq m_{i+1}$, $n_{i} \neq n_{i+1}$.
In the rest of the paper, with a slight abuse of notation, we refer to cycles in $\mathcal{G}(\HH)$ and cycles in $\HH$ interchangeably. To achieve a certain girth $g$, for given values of $m$ and $n$, and for a fixed value of $N$, one has to find a matrix $\mathbf{P}$ whose entries do not satisfy (\ref{fore}) for any value of $k < g/2$, and any possible choice of the row and column indexes $m_i$ and $n_i$.


\subsection{QC-SC-LDPCCs based on QC-LDPC codes \label{subsec:esdef} }


The edge spreading procedure \cite{Mitchell2017,Esfahanizadeh2017} is defined by an $m \times n$ $(m_s+1)$-ary \textit{spreading matrix} $\mathbf{B}$, where $m_s$ represents the \emph{memory} of the resulting SC-LDPCC. The spreading matrix $\BB$ can also be represented as a vector $\bb$ of length $n$, from which $\BB$ can be obtained by replacing each entry with the
associated $(m_s+1)$-ary column vector. A straightforward conversion from $\BB$ to $\bb$ is shown in Example \ref{exa:spre}. A \textit{convolutional exponent matrix} has the following form

\[ \mathbf{P}_{[0,\infty]} = \left[\begin{matrix}
 \mathbf{P}_{0}\\
  \mathbf{P}_{1}& \mathbf{P}_{0}\\
 \vdots&\mathbf{P}_{1}&\ddots\\
 \mathbf{P}_{m_s}&\vdots&\ddots \\
 & \mathbf{P}_{m_s}&\ddots\\
\end{matrix}\right],\]
where the $(i,j)$th entry of the $m\times n$ matrix  $\mathbf{P}_k$, $k\in[0,1,\ldots, m_s]$ is

\[
\mathbf{P}_k^{(i,j)}=\delta_k^{(i,j)}p_{i,j},
\]
where

\[
\delta_k^{(i,j)}=\begin{cases}
1 \quad \mathrm{if} \quad B_{i,j}=k\\
-\infty \quad \mathrm{if} \quad B_{i,j}\neq k,
\end{cases}
\]
and $B_{i,j}$ is the $(i, j)$th entry of $\mathbf{B}$.
Let us remark that $-\infty$ represents void entries in the convolutional exponent matrix and corresponds to the $N\times N$ all-zero matrix in the corresponding binary parity-check matrix. Notice that the entries of $\mathbf{P}_{[0,\infty]}$ which are off the main diagonal are $-\infty$ and have been omitted for the sake of readibility. The parity-check matrix of the QC-SC-LDPCC is then obtained as
\begin{equation}
\mathbf{H}_{[0,\infty]} = \left[\begin{matrix}
 \mathbf{H}_{0}\\
  \mathbf{H}_{1}& \mathbf{H}_{0}\\
 \vdots&\mathbf{H}_{1}&\ddots\\
 \mathbf{H}_{m_s}&\vdots&\ddots \\
 & \mathbf{H}_{m_s}&\ddots\\
\end{matrix}\right],
\label{eq:HSCQC}
\end{equation}
\noindent where the appropriate $N\times N$ \ac{CPMs} are substituted for the entries of $\PP_{[0,\infty]}$ which have values in the set $\{ 0,1,\ldots, N-1 \}$, and the $N\times N$ all-zero matrix is substituted for the entries of $\PP_{[0,\infty]}$ which are $-\infty$. $\mathbf{H}_{[0,\mathcal{L}]}$ represents a terminated version of $\mathbf{H}_{[0,\infty]}$, obtained by considering the first $(\mathcal{L}+m_s)Nm$ rows and $\mathcal{L}Nn$ columns of the semi-infinite parity-check matrix. For the sake of readability, in the rest of the paper we refer to QC-SC-LDPCCs based on QC-LDPC codes as QC-SC codes.

\begin{Exa}
Consider the $(3,5)$-regular array \ac{LDPC} block code 
with the exponent matrix
\begin{equation}
\PP = \left[\begin{matrix}
 0&0&0&0&0\\
 0&1&2&3&4\\
 0&2&4&1&3\\
\end{matrix}\right]
\label{eq:exarr5}
\end{equation}
and $N=5$. Consider also the spreading matrix, with $m_s = 2$,
\begin{equation}
\BB = \left[\begin{matrix}
 0&0&0&2&1\\
 0&1&2&1&0\\
 1&0&0&0&1\\
\end{matrix}\right],
\bb = \left[\begin{matrix}
1&3&6&21&10
\end{matrix}\right].
\label{eq:sprearr5}
\end{equation}
Then the constituent blocks of $\PP$ are
\[
\PP_0 = \left[\begin{matrix}
 0&0&0&-&-\\
 0&-&-&-&4\\
 -&2&4&1&-\\
\end{matrix}\right],
\PP_1 = \left[\begin{matrix}
 -&-&-&-&0\\
 -&1&-&3&-\\
 0&-&-&-&3\\
\end{matrix}\right],
\]
\[
\PP_2 = \left[\begin{matrix}
 -&-&-&0&-\\
 -&-&2&-&-\\
 -&-&-&-&-\\
\end{matrix}\right],
\]
where, for simplicity, $-\infty$ has been expressed as $-$.
\label{exa:spre}
\end{Exa}

\subsection{Exhaustive Search}

According to the definition given in Section \ref{subsec:esdef}, there are $(m_s+1)^{mn}$ possible spreading matrices. Nevertheless, some of them define equivalent codes. The size of the search space can be reduced, without loss of exhaustiveness, using the following property from \cite{Battaglioni2018}.

\begin{Lem}
Let $\PP_1$ and $\PP_2$ be exponent matrices. If $\PP_1$ can be obtained by permuting the rows or the columns of $\PP_2$, or if $\PP_1$ can be obtained by adding or subtracting (modulo $N$) the same constant to all the elements of a row or a column of $\PP_2$, then the corresponding codes are equivalent.
\label{lem:permequi}
\end{Lem}

It follows from Lemma \ref{lem:permequi} that the set of exponent matrices that contain at least one zero in each column represent, without loss of generality, the entire space of exponent matrices. Similarly, it is straightforward to show that the set of spreading matrices containing at least one zero in each column represent, without loss of generality, the entire space of spreading matrices. Each of the $m$ entries of a column of $\BB$ can assume values in $[0,1,\ldots,m_s]$ and, thus, there are $(m_s+1)^m$ possible columns. However, we can remove the $m_s^m$ columns which do not contain any zero. It follows that
\begin{equation}
[(m_s+1)^m-m_s^m]^n
\label{eq:candcol}
\end{equation}
spreading matrices cover the whole search space. It is straightforward to notice from \eqref{eq:candcol} that the number of candidate edge spreading matrices becomes very large as the values of $m$, $n$ and $m_s$ increase. For this reason, we propose, in Section \ref{sec:algo}, a novel procedure which allows distinguishing ``good'' candidates from ``bad'' candidates. Such an algorithm, based on a \textit{tree-search}, does not exclude, a priori, any candidate spreading matrix. Instead, ``bad'' candidates and their children are discarded by the algorithm during the search. In other words, the  algorithm only keeps ``good'' candidates, under the empirical assumption that the children of ``bad'' candidates are more likely to yield a higher multiplicity of harmful objects with respect to the children of ``good'' candidates. Numerical results provided in Section \ref{sec:perf} confirm that the aforementioned assumption is reasonable, since the proposed algorithm outputs spreading matrices yielding a smaller multiplicity of harmful objects with respect to previous approaches.

\subsection{Prior Work}

Previous works have also addressed the problem of reducing the search space of candidate spreading matrices. The most basic approach was proposed in \cite{Amiri2016}, where the authors minimize the number of $(3,3)$ \acp{AS} in $(3,n)$-regular array-based \acp{SC-LDPC-CC}, obtained through cutting vectors, which are a subclass of spreading matrices (see \cite{Mitchell2014} for further details). Such an approach is very efficient, since it relies on an integer optimization procedure, but the spanned search space is very small. Nevertheless, the cutting vectors, as defined in \cite{Amiri2016}, only permit to design \acp{SC-LDPC-CC} with memory $m_s=1$ and they only cover $\binom{n}{3}$ spreading matrices, instead of the total $8^n$ ($7^n$ with the reduction given by \eqref{eq:candcol}). This yields a non negligible chance that some optimal matrices are left out of the search.

In \cite{Mitchell2017} a guided random search is used to find optimal spreading matrices of $(3,n)$ regular array-based \acp{SC-LDPC-CC}, where a small subset of all the possible columns is considered, in such a way that the spreading matrix is ``balanced''. Although this approach can result in a quite fast search, especially if the subset contains a small number of elements, it is expected to be suboptimal, in that it a spans a search space which is considerably smaller than the whole one, without considering any optimization criterion. In particular, when $m_s=1$ (respectively, $m_s=2$), given that $m=3$, the guided random search in \cite{Mitchell2017} includes $5^n$ out of the total $8^n$ ($27^n$, respectively), possible spreading matrices, which can be reduced to $7^n$ ($19^n$, respectively), without loss of generality, according to Lemma \ref{lem:permequi}.

The method proposed in \cite{Esfahanizadeh2017} is similar to that proposed in \cite{Mitchell2017}. In fact, only a subset of all the possible spreading matrices is considered, such that each row contains $\frac{n}{m+1}$ entries\footnote{Approximation to the nearest integer is required when $m+1$ does not divide $n$.}  with value $i$, $0\leq i\leq m$. This also results in a sort of balanced spreading matrix. Nevertheless, also in this case, the search may not be optimal, since a large number of spreading matrices is excluded a priori (the exact number of candidates results in a long formula, which is omitted for space reasons, see \cite{Esfahanizadeh2017} for more details).

Finally, the approach in \cite{Beemer2017} relies on a searching algorithm which is not described in the original paper. For this reason, we are not able to estimate the number of candidates it considers. Nevertheless, in \cite{Beemer2017}, the authors mention that the search is limited; so, we conjecture that it suffers from the same problems of the methods proposed in \cite{Mitchell2017,Esfahanizadeh2017}.

\section{Edge spreading matrices \label{sec:edgesp}}

As mentioned in Section \ref{sec:TICodes}, trapping sets (and therefore absorbing and fully absorbing sets) originate from cycles, or clusters of cycles. In this section we prove conditions on the existence of cycles in $\HH_{[0,\infty]}$; this allows us to derive the number of equations that must be checked for each candidate spreading matrix in order to verify if it is a ``good'' candidate or a ``bad'' candidate for the proposed algorithm. The ``goodness'' of a candidate is measured by the number of harmful objects of the underlying block code it can eliminate.

We say that a \textit{block-cycle} with length $\lambda$ exists in the Tanner graph corresponding to the parity-check matrix of the block code described by $\PP$ if there exists an $m \times n$ submatrix of $\PP$, denoted as $\PP^{\lambda}$, containing $\lambda$ of its non-void entries (and $-\infty$ elsewhere) such that \eqref{fore} holds.

 The \textit{block-cycle distribution} (or \textit{spectrum}) of $\HH_{[0,\mathcal{L}]}$ is denoted as $\mathbf{D}^{\mathcal{L},\Lambda}$ and is a vector such that its $i$th entry $\mathbf{D}_i^{\mathcal{L},\Lambda}$ represents the multiplicity of block-cycles with length $2i+4\leq \Lambda$ in $\mathcal{G}(\HH_{[0,\mathcal{L}]})$.

We calculate the \textit{average number of block-cycles with length $\lambda$ per node} $E_{\lambda}$ as follows:

\begin{enumerate}
\item evaluate the number of block-cycles spanning exactly $i$ sections, $i \in [2,3,\ldots, \lfloor\frac{\lambda}{4}\rfloor m_s+1]$ as
\begin{equation}
    K_i=D^{i,\lambda}_{\frac{\lambda-4}{2}}-\sum_{j=1}^{i-1}(i+1-j)K_j,
    \label{eq:exaspancy}
\end{equation}
where $K_1 = D^{1,\lambda}_{\frac{\lambda-4}{2}}$;
\item compute the average as 
\begin{equation}
    E_{\lambda}=\frac{\sum_{i=1}^{\lfloor\frac{\lambda}{4}\rfloor m_s+1} K_i}{n}.
    \label{eq:avg}
\end{equation}
\end{enumerate}

We also define $\mathbf{E}^{\Lambda}$ as the vector containing $E_{\lambda}$, $\forall \lambda \in [4,6,\ldots,\Lambda]$, as its entries. A similar procedure can be used to compute the average number of $(a,b)$ absorbing sets, $E_{(a,b)}$. 

The following result holds.
\begin{Lem}
Consider a \textit{block-cycle} with length $\lambda$, described by $\PP^{\lambda}$, existing in the Tanner graph $\mathcal{G}(\HH)$ corresponding to the parity-check matrix of the block QC-LDPC code described by $\PP$. Then, after the edge spreading procedure based on $\BB$ is applied, such a block-cycle also exists in $\mathcal{G}(\HH_{[0,\infty]})$ if and only if $\BB^{\lambda}$ satisfies \eqref{fore} over $\mathbb{Z}$, where
\[\begin{cases}
B^{\lambda}_{i,j} = -\infty \quad \mathrm{if} \quad P^{\lambda}_{i,j}=-\infty,\\
B^{\lambda}_{i,j}=B_{i,j} \quad \mathrm{otherwise}.
\end{cases}\]
\label{lem:bandpfosso}
\end{Lem}

\begin{IEEEproof}
Let us derive from $\PP$ a matrix $\mathbf{R}$  as follows
\[\begin{cases}
R_{i,j} = 0 \quad \mathrm{if} \quad P_{i,j}=-\infty,\\
R_{i,j} = 1  \quad \mathrm{otherwise}.
\end{cases}\]
Suppose that a simple cycle $\mathcal{C}$ with length $\lambda$ exists in $\mathcal{G}(\mathbf{R})$. The spreading operation defined by $\BB$ yields a matrix $\mathbf{R}_{[0,\infty]}$ such that $\mathcal{G}(\mathbf{R}_{[0,\infty]})$ will still contain $\mathcal{C}$ if and only if the entries of $\BB$ that are in the same positions as the $1$s involved in the cycle satisfy \eqref{fore} over $\mathbb{Z}$. It is clear that any block-cycle in $\mathcal{G}(\mathbf{H}_{[0,\infty]})$ corresponds to a simple cycle in $\mathcal{G}(\mathbf{R}_{[0,\infty]})$ (however the converse, in general, is not true). Since we assumed that $\PP^{\lambda}$ describes a block-cycle with length $\lambda$, $\mathcal{G}(\mathbf{H}_{[0,\infty]})$ will also contain this block-cycle if and only if the $\lambda$ entries of $\BB$ that are in the same positions as the $\lambda$ entries of $\PP^{\lambda}$ that are not $-\infty$ satisfy \eqref{fore} over $\mathbb{Z}$. 
\end{IEEEproof}

Suppose now that the code defined by an exponent matrix $\PP$ contains $\nu$ block-cycles. Given $\BB$, we can extract all the  submatrices $\BB^{\lambda_i}$, $0 \leq i \leq \nu-1$, that correspond to the block-cycles in the QC-LDPC code and check whether \eqref{fore} is satisfied. If it is, then the block-cycle also exists in the QC-SC code; if it is not, then the block-cycle does not exist in the QC-SC code. In other words, given an exponent matrix and a spreading matrix, checking as many equations as the number of block-cycles in the exponent matrix will determine the number of block-cycles in the convolutional exponent matrix. We also remark that a block-cycle in an exponent matrix corresponds to $N$ cycles in the binary parity-check matrix.

\begin{Exa}
Consider the same code and the same spreading matrix as in Example \ref{exa:spre} (see \eqref{eq:exarr5} and \eqref{eq:sprearr5}, respectively). $\mathcal{G}(\HH)$ contains twenty block-cycles with length $\lambda=6$. For the sake of brevity, we only consider three of them, along with the corresponding entries of the spreading matrix
\[
\PP^{\lambda_0} = \left[\begin{matrix}
 0&0&-&-&-\\
 0&-&2&-&-\\
 -&2&4&-&-\\
\end{matrix}\right] \BB^{\lambda_0} = \left[\begin{matrix}
 0&0&-&-&-\\
 0&-&2&-&-\\
 -&0&0&-&-\\
\end{matrix}\right],
\]

\[
\PP^{\lambda_1} = \left[\begin{matrix}
 -&0&0&-&-\\
 0&-&2&-&-\\
 0&2&-&-&-\\
\end{matrix}\right] \BB^{\lambda_1} = \left[\begin{matrix}
 -&0&0&-&-\\
 0&-&2&-&-\\
 1&0&-&-&-\\
\end{matrix}\right],
\]

\[
\PP^{\lambda_2} = \left[\begin{matrix}
 -&0&0&-&-&\\
 -&1&-&3&-\\
 -&-&4&1&-\\
\end{matrix}\right] \BB^{\lambda_2} = \left[\begin{matrix}
 -&0&0&-&-\\
 -&1&-&1&-\\
 -&-&0&0&-\\
\end{matrix}\right]
\]

Notice that $\PP^{\lambda_i}$, $i=0,1,2$, comply with \eqref{fore}, as they represent block-cycles in the array LDPC block code. Moreover, \eqref{fore} is satisfied for $\BB^{\lambda_2}$ but not for $\BB^{\lambda_0}$, $\BB^{\lambda_1}$. In other words, $\mathcal{G}(\HH_{[0,\infty]})$ contains the block-cycles of length $6$ corresponding to $\PP^{\lambda_2}$, but not those associated to $\PP^{\lambda_0}$ and $\PP^{\lambda_1}$. The same procedure can be applied to test whether the remaining $17$ block-cycles are also contained in $\mathcal{G}(\HH_{[0,\infty]})$ or not.
\label{exa:exbpfosso}
\end{Exa}

\section{A Greedy Algorithm to Construct Optimized QC-SC Codes} \label{sec:algo}

\begin{algorithm}[th!]
\caption{\label{alg:pseudo}}
\textbf{Input} exponent matrix $\PP$, circulant size $N$, size of harmful objects $\lambda$, all-zero spreading matrix $\BB$, memory $m_s$ \\
\begin{algorithmic}
\Procedure {\texttt{mihao}}{$\PP$, $N$, $\lambda$, $\BB$, $m_s$}
\State {$\BB_{\mathrm{old}} \gets \BB$}
\State{$\HH \gets \mathrm{edge\_spread}(\PP,\BB,N)$}
\State {$C_{\mathrm{old}} \gets \mathrm{count\_harmful\_objects}(\HH, \lambda)$}
\For{$i \gets 0 \lr m$}
\For{$j \gets 0 \lr n$}
\If{$\BB_{i,j}=0$}
\For{$k \gets 0 \lr m_s$}
\State{$\BB_{i,j}\gets k$}
\State{$\mathbf{M}^{(k)}_{i,j}\gets \mathrm{count\_elimin\_objects}(\PP,\BB)$}
\State{$\BB_{i,j}\gets 0$}
\EndFor
\EndIf
\EndFor
\EndFor
\State{$M \gets \max_{0\leq k \leq m_s}\mathbf{M}^{(k)}_{i,j} $}
\State{$n_{\mathrm{cands}} \gets \#(\mathbf{M}^{(k)}_{i,j}=M) $}
\While{!Stopping criterion}
\If {$n_{\mathrm{cands}}>0$}

\State{Randomly pick $(i,j,k)$ such that $\mathbf{M}^{(k)}_{i,j}=M $}
\State {$\BB_{\mathrm{new}} \gets \BB$}
\State{$\BB_{\mathrm{new}}^{(i,j)}\gets k$}
\State{$\HH \gets \mathrm{edge\_spread}(\PP,\BB_{\mathrm{new}},N)$}
\State {$C_{\mathrm{new}} \gets \mathrm{count\_harmful\_objects}(\HH, \lambda)$}

\If{$C_{\mathrm{new}}<C_{\mathrm{old}}$}
\State{$\BB \gets$ \texttt{MIHAO}($\PP$, $N$, $\lambda$, $\BB_{\mathrm{new}}$, $m_s$)}
\Else
\State{$\BB\gets\BB_{\mathrm{old}}$}
\EndIf

\State{$n_\mathrm{cands} \gets n_\mathrm{cands} - 1$}
\State{$\mathbf{M}_{i,j}^{(k)}\gets 0$}

\Else
\State{$\BB_{\mathrm{out}}\gets\BB_{\mathrm{old}}$}
\State{\Return{$\BB_{\mathrm{out}}$}}
\EndIf
\EndWhile
\EndProcedure
\end{algorithmic}
\end{algorithm}

In this section we describe a general algorithm, named MInimization of HArmful Objects (MIHAO), which can be applied to an arbitrary harmful object (or objects) of interest to find a good QC-SC code. Given the exponent matrix of a QC-LDPC block code, we first determine which are the most harmful objects causing an error rate performance degradation. The pseudo-code describing the proposed recursive procedure is described in Algorithm \ref{alg:pseudo}.

We propose to use a tree-based search: the root node of the tree is the all-zero spreading matrix, which characterizes a QC-LDPC block code; the $l$th tier contains all the spreading matrices with $l$ non-zero entries which minimize the multiplicity of harmful objects with respect to their parent node. If a parent node has no children nodes with better properties than its own, it is discarded, and the algorithm backtracks. If no specific stopping criterion is included, all the candidates are tested; the node representing the spreading matrix yielding the smallest number of harmful objects is the output of the algorithm. Stopping criteria can be, for example, the maximum number of times the algorithm backtracks or the maximum number of tiers it spans.

In particular, we provide in the following a description of the functions used throughout Algorithm \ref{alg:pseudo}. The function {\fontfamily{cmss}\selectfont edge\_spread$(\PP,\BB,N)$} performs the edge spreading procedure as described in Section \ref{subsec:esdef}; {\fontfamily{cmss}\selectfont count\_elimin\_objects$(\PP,\BB)$} determines how many harmful objects are removed from $\PP$ for a given $\BB$. This is accomplished according to Remark \ref{lem:bandpfosso}, as shown in Example \ref{exa:exbpfosso}. Then, the candidate base matrices are those maximizing the multiplicity of removed harmful objects. Finally, {\fontfamily{cmss}\selectfont count\_harmful\_objects$(\HH,\lambda)$} computes the multiplicity of harmful objects of length $\lambda$ in $\HH$. This function is inspired by the counting algorithm proposed in \cite{Zhou2010}. The metric we finally consider to determine whether the candidate is ``good'' or ``bad'' is the average number of harmful objects per node, as defined in Section \ref{sec:edgesp}.

Note that the algorithm does not guarantee that the optimal solution, which is obviously unknown, will be the output but, as will be shown in  Section \ref{sec:perf}, it provides better solutions than the best available in the literature. 

\section{Numerical Results and Performance \label{sec:perf}}

We validate the procedure using array codes \cite{Fan2000} and Tanner codes \cite{Tanner2004} as a benchmark; then, confirm the expected performance improvement via Monte Carlo simulations.

\subsection{Optimization results \label{subsec:tabless}}

It is known that the performance of $(3,n)$-regular array codes is adversely affected by $(3,3)$ \acp{AS} and $(4,2)$ FASs. It can be easily shown that $(3,3)$ \acp{AS} and $(4,2)$ FASs derive from a cycle with length $6$ and a cluster of two cycles with length $6$, respectively \cite{Mitchell2014}. We have applied Algorithm 1 to minimize their multiplicity in array-based QC-SC codes when $m_s=1$. The results are shown in Table \ref{table:Tabprev}.
\begin{table}[!t]
\caption{Average number of $(3,3)$ absorbing sets per node $E_{(3,3)}$ in array-based SC-LDPC codes with $m=3$, $m_s=1$}
\label{table:Tabprev}
\centering
\begin{tabular}{|c|c|c|c|c|c|c|c|c|c|c|c|c|c|c|c|}
\hline
$p$  & {7} & {11} & {13} &{17}&
{19}&
{23}\\ \hline \hline
$E_{(3,3)}$   & 0.43  &1  & $1.08$ &  $1.88$ & $2.26$&$3.26$ \\ \hline
$E_{(3,3)}$ Literature  & 0.43 & 1 & $1.23$ & $1.88$& $2.68$& $3.78$\\ \hline
\end{tabular}
\end{table} 

We have also considered the $(3,5)$-regular Tanner QC-LDPC code with $L=155$ and $g = 8$, described by

\begin{equation}
\PP_{\frac{2}{5}}=\left[ \begin{matrix}
1 & 2& 4&8&16\\
5&10&20&9&18\\
25&19&7&14&28\\
\end{matrix}\right].
\label{eq:Ptanner}
\end{equation}

The dominant trapping sets of this code are known to be $(8,2)$ \acp{AS} \cite{Zhang2011}. They consist of clusters of cycles with length $8$, $10$, $12$,  $14$ and $16$. The easiest approach to eliminate these sets is to target the shortest cycles for removal. By applying Algorithm 1 with the following inputs: $\PP_{\frac{2}{5}}$, $N=31$, $\lambda=8$, the all-zero spreading matrix $\BB$, and $m_s=1$, we obtain 
\begin{equation}
\bb_1=\left[ \begin{matrix}
2&2&1&1&4
\end{matrix}\right],
\label{eq:Btanner}
\end{equation}
which results in a QC-SC parity-check matrix with no cycles of length up to $8$. We have $\mathbf{E}^{12}=\left[\begin{matrix}
0 & 0 & 0 & 3.8 & 18.4\\
\end{matrix}\right].$
One can also minimize the multiplicity of cycles of length $10$ and $12$, by applying Algorithm \ref{alg:pseudo} with different values of $\lambda$.
For $g=10$ and $\lambda=12$, we obtained
\begin{equation}
\bb_2=\left[ \begin{matrix}
2 &1 &6 &1 &5
\end{matrix}\right],
\label{eq:Btannernew}
\end{equation}
where
$\mathbf{E}^{12}=\left[\begin{matrix}
0 & 0 & 0 & 1.8 & 15
\end{matrix}\right].$
Further improvement can be obtained by applying Algorithm 1 to eliminate all the block-cycles with length $10$. This requires an increase in the memory to $m_s=3$ and results in the spreading matrix
\begin{equation}
\bb_3=\left[ \begin{matrix}
35&12&50&50&15
\end{matrix}\right],
\label{eq:Btannermh3}
\end{equation}
which yields $\mathbf{E}^{12}=\left[\begin{matrix}
0 & 0 & 0 & 0 & 9.4
\end{matrix}\right].
$ Note that an exhaustive search for such a code demands a huge computational effort, since it would require to perform $69343957$ attempts.

Suppose we wish to reduce the multiplicity of cycles with length $12$, which are known to combine to create codewords of minimum weight 24. From the exponent matrix \eqref{eq:Ptanner}, Algorithm 1 with $m_s=1$ outputs the edge-spreading matrix
\begin{equation}
\bb_4=\left[ \begin{matrix}
6&1&3&2&4
\end{matrix}\right]
\label{eq:Btanner5few12}.
\end{equation}
In this case we have $\mathbf{E}^{12}=
\left[\begin{matrix}
0 & 0 & 0.6 & 3.2 & 14.2
\end{matrix}\right]
$.

As a final example, we consider the $(3,7)$-regular Tanner code with blocklength $L=301$, $g=8$ and
\begin{equation}
\PP_{\frac{4}{7}}=\left[ \begin{matrix}
1    & 4 &   16 &   21 &   41  &  35  &  11\\
     6  &  24   & 10   & 40  &  31   & 38   & 23\\
    36   & 15  &  17  &  25  &  14  &  13 &    9\\
\end{matrix}
\right],
\label{eq:tan7}
\end{equation}
from which two QC-SC codes have been obtained with spreading matrices

\begin{equation}
\bb_5=\left[ \begin{matrix}
3&4&2&4&1&6&6
\end{matrix}\right]
\label{eq:Btanner7alot12},
\end{equation}
\begin{equation}
\bb_6=\left[ \begin{matrix}
5&3&1&4&6&2&4
\end{matrix}\right]
\label{eq:Btanner7few12}.
\end{equation}

Matrix $\bb_5$ was randomly generated with $m_s=1$, whereas $\bb_6$ is the output of Algorithm \ref{alg:pseudo} with inputs $\PP_{\frac{4}{7}}$, $N=43$, $\lambda=12$, the all-zero spreading matrix $\BB$, and $m_s=1$. The respective block-cycle distributions of these two codes are \[\mathbf{E}^{12}=\left[\begin{matrix}
0 & 0 & 1.86 &  17.57  & 71.14
\end{matrix}\right],\]
\[\mathbf{E}^{12}=\left[\begin{matrix}
0 & 0 & 1.29 & 15.14 &   64 \\
\end{matrix}\right].\]

We have compared the time taken by Algorithm \ref{alg:pseudo} to output all these spreading matrices with the average time required to find spreading matrices with the same (or better) cycle spectra through random searches. The average speed up obtained is shown in Table \ref{table:Tabtime}, where $t_{\mathrm{ran}}$ and $t_{\mathrm{alg}}$ are the times required by the random search and by Algorithm \ref{alg:pseudo}, respectively.

\begin{table}[!t]
\caption{Average speed up of Algorithm \ref{alg:pseudo} with respect to random search}
\label{table:Tabtime}
\centering
\begin{tabular}{|c|c|c|c|c|c|c|c|c|c|c|c|c|c|c|c|}
\hline
Code  & $\BB_1$ & $\BB_2$ & $\BB_3$ & $\BB_4$ &
$\BB_6$\\ \hline \hline
$\frac{t_{\mathrm{ran}}}{t_{\mathrm{alg}}}$ & $3.73$  & $4.2$  & $8.21$ &  $3.51$ & $4.18$\\ \hline
\end{tabular}
\end{table}

\subsection{Monte Carlo simulations}
In this section we assess the performance of the newly designed codes described in Section \ref{subsec:tabless} in terms of \ac{BER} via Monte Carlo simulations of \ac{BPSK} modulated transmissions over the \ac{AWGN} channel. We have used a \ac{SW} decoder with window size (in periods) $W=5(m_s+1)$ performing $100$ iterations. The \ac{SW} decoder performs belief propagation over a window including $W$ blocks of $L$ bits each, and then let this window slide forward by $L$ bits before starting over again. For each decoding window position, the SW decoder gives the first $L$ decoded bits, usually called \textit{target bits}, as output.

\begin{figure}
\begin{center}
\includegraphics[width=85mm,keepaspectratio]{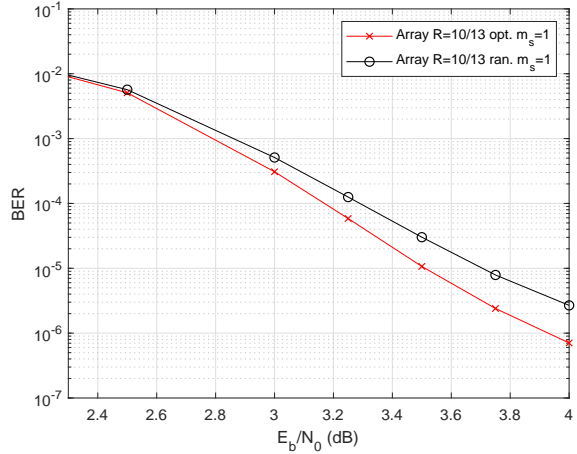}
\caption{Simulated performance of array-based SC codes as a function of the signal-to-noise ratio.\label{fig:perfarr}}
\end{center}
\end{figure}

First, we have considered the $(3,13)$-regular array code and we have simulated the QC-SC code obtained by edge-spreading its exponent matrix $\PP$ with the optimized spreading matrix found by Algorithm \ref{alg:pseudo} (the number of harmful objects is given in Table \ref{table:Tabprev}) and with a random spreading matrix. The results shown in Fig. \ref{fig:perfarr} confirm that $(3,3)$ absorbing sets have a significant impact on these codes and enforce the necessity of an effective design to reduce their multiplicity. 

\begin{figure}
\begin{center}
\includegraphics[width=85mm,keepaspectratio]{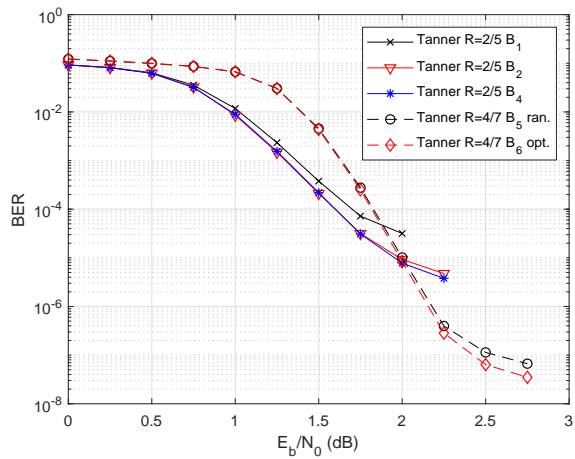}
\caption{Simulated performance of  Tanner-based SC codes as a function of the signal-to-noise ratio.\label{fig:perfTan}}
\end{center}
\end{figure}

We have also considered the $(3,5)$-regular Tanner code and simulated the QC-SC codes obtained by edge-spreading \eqref{eq:Ptanner} with $\BB_1$ and $\BB_2$. The results, shown in Fig. \ref{fig:perfTan}, confirm the effectiveness of Algorithm 1. We have also analyzed the decoding failure patterns of these codes and noticed that, according to the analysis proposed in \cite{Battaglioni2018a}, many of them were caused by cycles of length $12$. 
For this reason, we have simulated the QC-SC code represented by $\BB_4$.
It can be noticed that, even though $\mathcal{G}(\HH_{[0,\infty]})$ for \eqref{eq:Btanner5few12}
contains some block-cycles with length $8$ and $10$, there is an improvement due to the reduction of the multiplicity of block-cycles with length $12$. The same approach has been followed for the QC-SC codes represented by \eqref{eq:Btanner7alot12} and \eqref{eq:Btanner7few12} ($\BB_5$ and $\BB_6$) that are constructed from the $(3,7)$-regular Tanner code.
According to their block-cycle spectra, the multiplicity of block-cycles with length $12$ was minimized for \eqref{eq:Btanner7few12}. This is seen to have a positive impact on the \ac{BER} performance in Fig.~\ref{fig:perfTan}.

\section{Conclusion \label{sec:conc}}

We have proposed an efficient algorithm enabling optimization of QC-SC codes based on QC-LDPC block codes from the perspective of harmful objects. The algorithm is flexible and allows the analysis of codes with different structure and values of memory and rate. Many classes of harmful objects can be the target of a search-and-remove process aimed at optimizing codes in terms of error rate performance.

\section*{Acknowledgement}

This material is based upon work supported by the National Science Foundation under Grant No. ECCS-1710920.

\bibliographystyle{IEEEtran}
\bibliography{Archive}

\end{document}